# Implementation of a Hybrid Scheme Combining Post-Quantum Cryptography and Quantum Key Distribution: A Case Study of ETSI Specifications

Abel C. H. Chen
*Information & Communications Security Laboratory,*
*Chunghwa Telecom Laboratories*
Taoyuan, Taiwan
Email: chchen.scholar@gmail.com; ORCID: 0000-0003-3628-3033

***Abstract*—With the increasing maturity of quantum computing and quantum communication technologies, Quantum Key Distribution (QKD) has gradually become an important approach for establishing shared secret keys between communicating parties. The European Telecommunications Standards Institute (ETSI) QKD Working Group has developed a series of group specifications (GSs) for QKD technologies. For instance, ETSI GS QKD 014 specifies the communication interface between the Secure Application Entity (SAE) and the Key Management Entity (KME), providing a foundation for the implementation and interoperability of QKD-related communication interfaces in industry. However, although ETSI GS QKD 014 defines the relevant communication interface, the overall solution does not yet fully achieve a quantum-safe level of security. Therefore, this study proposes a quantum-safe QKD network architecture to establish quantum-safe communications between SAEs and KMEs. Furthermore, the study proposes a QKD-based Transport Layer Security (TLS) scheme and a hybrid TLS scheme integrating Elliptic Curve Cryptography (ECC), Post-Quantum Cryptography (PQC), and QKD. These approaches are intended to provide applications with a more secure communication foundation. Experimental evaluations are conducted to compare the proposed approaches with state-of-the-art (SOTA) methods from multiple perspectives, including security and message size. The results provide an assessment of the security characteristics and communication overhead of the proposed approaches in comparison with existing solutions.**

## I. INTRODUCTION

Following the developments of the BB84 protocol [1] and the E91 protocol [2], increasing attention has been devoted to Quantum Key Distribution (QKD) in the research community. Subsequent studies have investigated improvements to QKD protocols [3] as well as their applications in various fields [4]. Furthermore, with the advancement of quantum computing and quantum communication technologies, QKD-related technologies have gradually progressed from theoretical research toward practical implementations. Recent implementations have achieved improved performance and reduced error rates, facilitating their applications [5]. In recent years, the European Telecommunications Standards Institute (ETSI) QKD Working Group has developed a series of group specifications (GSs) addressing QKD network architectures and communication interfaces [6]-[8]. In particular:

Although ETSI has established GSs for various QKD communication interfaces, some interfaces do not yet provide a fully quantum-safe level of security and may therefore remain vulnerable to attacks enabled by sufficiently powerful quantum computers. For instance, the communication interface between an SAE and a KME may still rely on conventional cryptographic mechanisms, such as RSA or Elliptic Curve Cryptography (ECC), which are not considered quantum-safe against large-scale quantum computing attacks.

To address the aforementioned issues, this study proposes a quantum-safe QKD network architecture to provide quantum-safe mutual authentication and key exchange between Secure Application Entities (SAEs) and Key Management Entities (KMEs). In addition, a QKD-based Transport Layer Security (TLS) protocol (named “QKD-based TLS”) is proposed. To further enhance the security of the shared secret (SS), a hybrid TLS protocol integrating ECC, Post-Quantum Cryptography (PQC)(named “hybrid QKD-ECC-PQC-based TLS”), and QKD is also proposed. The main contributions of this study are summarized as follows:

- The study enhances the QKD network architecture specified by the ETSI and proposes a quantum-safe QKD network architecture to establish quantum-safe communications between SAEs and KMEs.
- The QKD-based TLS is proposed. The proposed protocol integrates the quantum-safe QKD network architecture to perform QKD and establish a SS between the communicating parties.
- The hybrid QKD-ECC-PQC-based TLS is proposed. The security of the resulting SS is based on multiple security foundations, including Quantum Property Preservation (QPP) [9], the Learning With Errors (LWE) problem [10], and the Elliptic Curve Discrete Logarithm Problem (ECDLP) [11].

This paper is organized into five sections. Section II reviews the state-of-the-art (SOTA) approaches related to ETSI QKD network architectures and TLS protocols. Section III presents the proposed quantum-safe QKD network architecture, QKD-based TLS, and hybrid QKD-ECC-PQC-based TLS. Section IV describes the evaluation of the proposed approaches and compares them with SOTA approaches in terms of security and data size. Finally, Section V concludes the paper by summarizing the main contributions and discussing potential directions for future research.

## II. LITERATURE REVIEWS

### *A. QKD Network Architecture Based on ETSI GSs*

The QKD network architecture and operational flow specified by the ETSI are illustrated in Fig. 1. The architecture primarily consists of SAEs and Trusted Nodes (TNs). Each TN includes a KME and a Quantum Key Distribution Entity (QKDE) [6]-[8]. In this study, a TN is assumed to operate similarly to a Hardware Security Module (HSM), where communications within the TN, i.e., between the KME and QKDE, are assumed to be established over a secure communication environment.

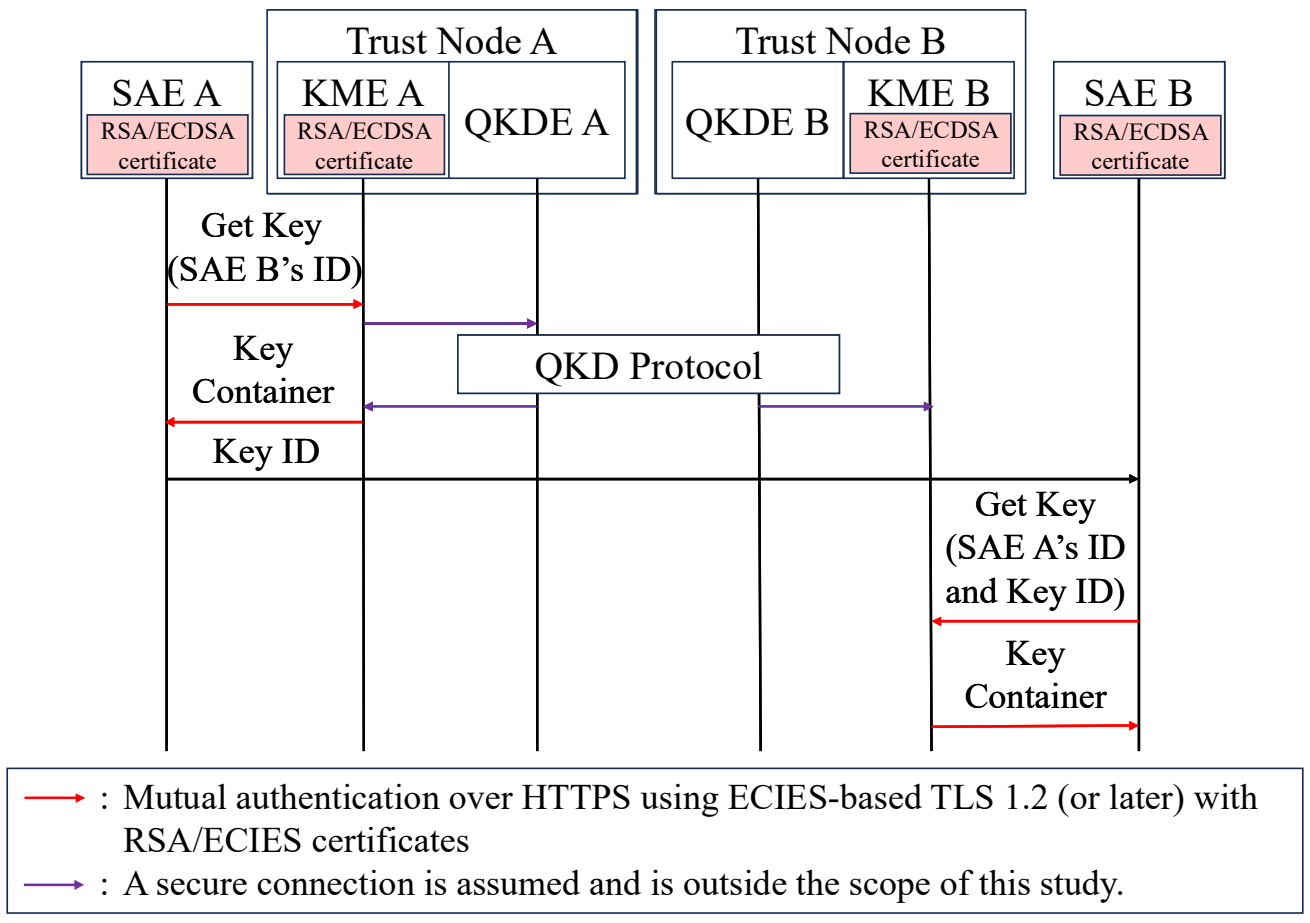

Fig. 1. The QKD network architecture based on ETSI GSs

To illustrate the QKD network architecture and operational procedures specified by the ETSI, the following example considers two SAEs and two TNs, namely SAE A, TN A, SAE B, and TN B. SAE A is connected to TN A, while SAE B is connected to TN B. Moreover, the QKDE A within TN A and QKDE B within TN B are assumed to communicate using a QKD protocol.

ETSI GS QKD 014 assumes that the communicating parties, represented by SAE A and SAE B in this case, already know each other's identities (IDs). Therefore, when the communicating parties need to obtain a SS, SAE A can invoke the Get Key interface of KME A within TN A and provide the ID of SAE B as the target SAE. Subsequently, KME A and KME B can communicate through the QKD protocol to obtain a SS. KME A then returns a Key Container object containing the Key ID and the SS to SAE A. SAE A subsequently provides the Key ID to SAE B. SAE B can then invoke the Get Key interface of KME B by providing the ID of SAE A and the Key ID, thereby obtaining a Key Container object containing the same key information.

However, ETSI GS QKD 014 was published in 2019 [7], at a time when the National Institute of Standards and Technology (NIST) had not yet established its PQC standards. Furthermore, the Internet Engineering Task Force (IETF) had not yet standardized PQC-based X.509 certificates or PQC-based TLS protocols. Consequently, the cryptographic mechanisms referenced and adopted by ETSI GS QKD 014 are primarily based on conventional cryptographic algorithms, including RSA and ECC, such as the Elliptic Curve Digital Signature Algorithm (ECDSA) and the Elliptic Curve Integrated Encryption Scheme (ECIES). Therefore, although mutual authentication between SAEs and KMEs can be achieved through certificates and digital signatures, and the communication channels can be established using TLS, the underlying cryptographic mechanisms are still based on RSA and ECC. These conventional public-key cryptographic mechanisms are not considered quantum-safe against sufficiently powerful quantum computers. Therefore, the components highlighted in red in Fig. 1 do not provide a quantum-safe security level and may potentially be vulnerable to attacks based on quantum computing.

## B. Transport Layer Security

### 1) ECC-based TLS

This section describes an ECC-based TLS protocol, where End Entity 1 (EE1) acts as the initiator and EE2 acts as the responder, as illustrated in Fig. 2. First, EE1 generates an ephemeral ECC key pair and sends a ClientHello message, in which the public key of the ephemeral ECC key pair is included in the key_share field. Upon receiving the ClientHello message, EE2 generates its own ephemeral ECC key pair and sends a ServerHello message, in which the corresponding ephemeral ECC public key is included in the key_share field [12]. The two communicating parties can then use their respective ephemeral ECC private keys and the peer's ephemeral ECC public key to perform Elliptic Curve Diffie-Hellman (ECDH) key exchange. The resulting SS, denoted as $r_E$, is treated as a shared pseudorandom value. Subsequently, $r_E$ is provided as input to a Key Derivation Function (KDF) to derive the shared key (SK).

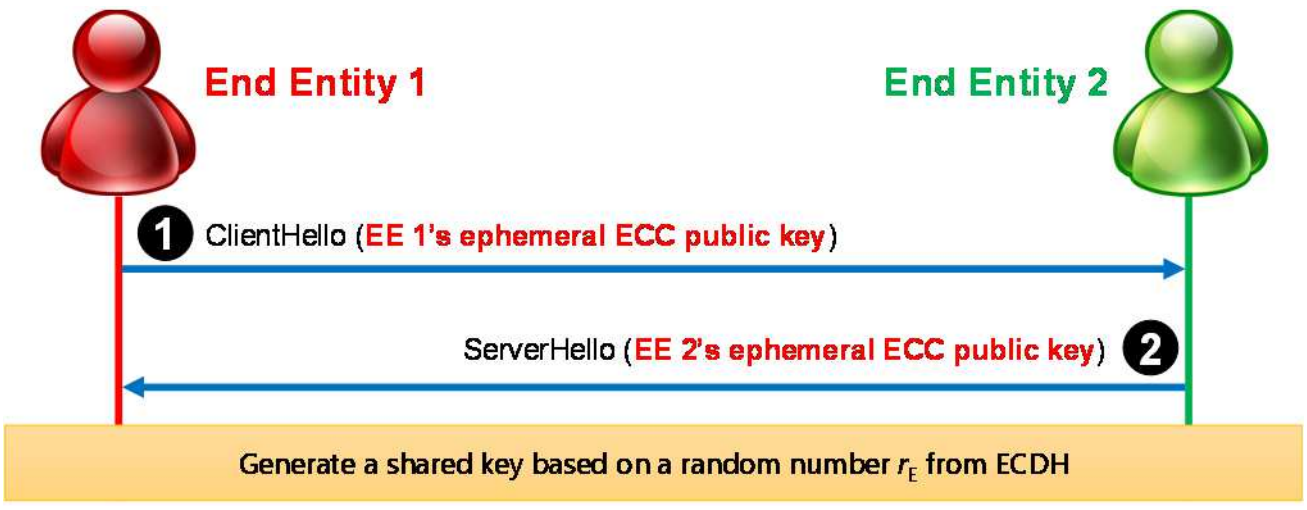

Fig. 2. ECC-based TLS

### 2) Hybrid ECC-PQC-based TLS

This section describes a hybrid TLS protocol based on ECC and PQC (named "hybrid ECC-PQC-based TLS"). In this scenario, EE1 acts as the initiator and EE2 acts as the responder. In addition to the ECC mechanism described in Section III.B.1, both parties employ a PQC mechanism, such as the Module-Lattice-Based Key-Encapsulation Mechanism (ML-KEM), as illustrated in Fig. 3. First, EE1 generates an ephemeral ECC key pair and an ephemeral ML-KEM key pair. EE1 then sends a ClientHello message, in which the public keys of the ephemeral ECC and ML-KEM key pairs are included in the key_share field. Upon receiving the ClientHello message, EE2 generates an ephemeral ECC key pair and uses EE1's ephemeral ML-KEM public key to perform the ML-KEM encapsulation operation. This operation generates an ML-KEM SS, denoted as $r_P$, and the encapsulation ciphertext. EE2 then sends a ServerHello message containing its ephemeral ECC public key and the ML-KEM encapsulation ciphertext in the key_share field [13]. Upon receiving the encapsulation ciphertext, EE1 uses its ephemeral ML-KEM private key to perform the ML-KEM decapsulation operation and recovers the same ML-KEM SS $r_P$. Subsequently, the two communicating parties use their respective ephemeral ECC private keys and the peer's ephemeral ECC public key to perform ECDH key exchange, thereby obtaining an ECC-based SS $r_E$, which is treated as a shared pseudorandom value. Finally, $r_E$ and $r_P$ are provided as inputs to a KDF to derive the SK. RFC 10024 specifies the key agreements and group IDs for hybrid schemes combining ECC and PQC, including X25519MLKEM768, SecP256r1MLKEM768, and SecP384r1MLKEM1024 [14].

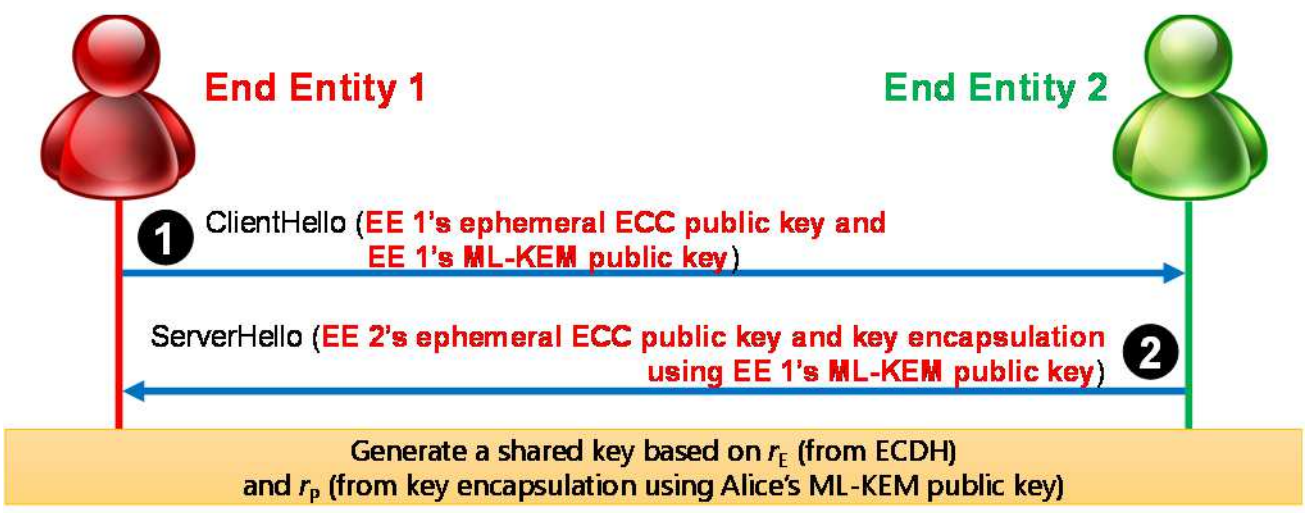

Fig. 3. Hybrid ECC-PQC-based TLS

## III. The Proposed Methods

### A. *The Proposed Quantum-Safe QKD Network Architecture*

To enhance the security of the QKD network architecture currently specified in ETSI GS QKD 014, this study proposes a quantum-safe QKD network architecture, as illustrated in Fig. 4. The main contributions are summarized as follows.

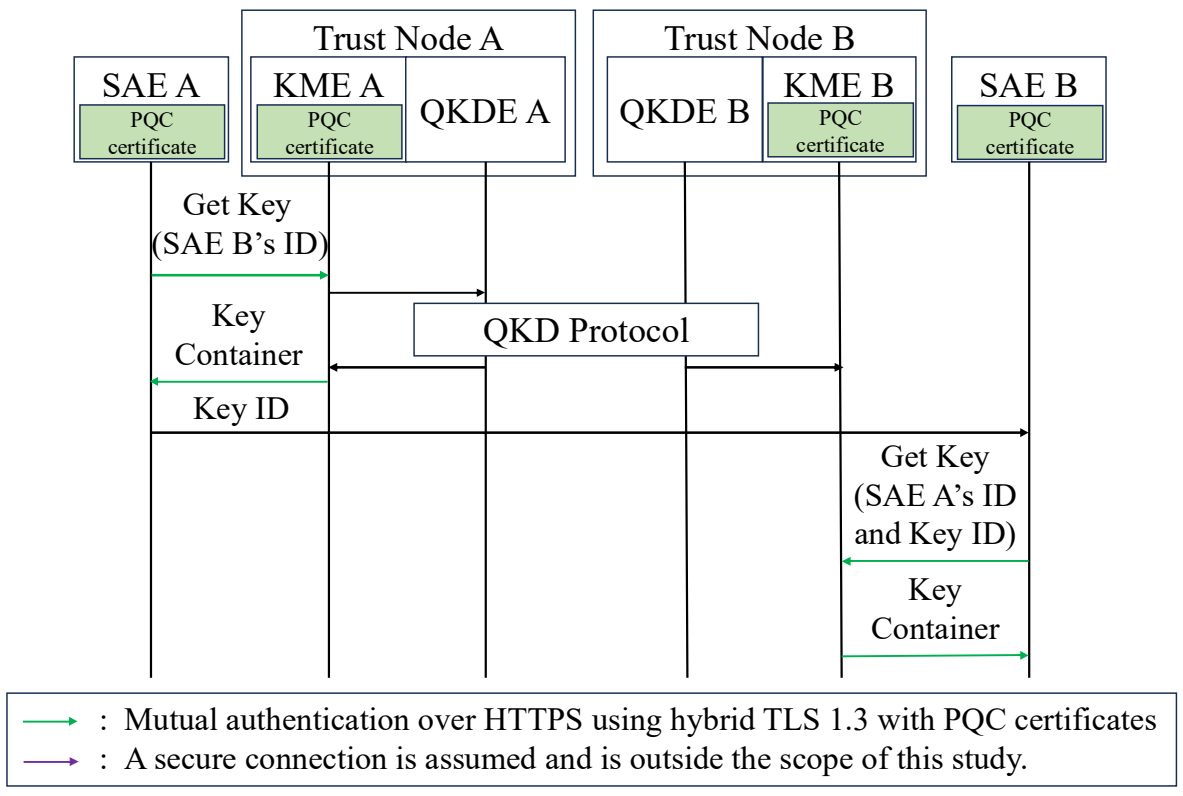

Fig. 4. The proposed quantum-safe QKD network architecture

- A trusted Certificate Authority (CA) issues a PQC-based X.509 certificate to each SAE and each KME [15]. The primary PQC digital signature algorithm considered in this study is the Module-Lattice-Based Digital Signature Algorithm (ML-DSA) [16]. The corresponding X.509 certificate formats are specified in RFC 9881 [17] and RFC 9882 [18]. In the proposed architecture, it is recommended that the Common Name (CN) field in each certificate be consistent with the corresponding SAE ID or KME ID, thereby providing an additional basis for subsequent device identity authentication.
- The hybrid ECC-PQC key exchange mechanism [13] is adopted to enable the establishment of quantum-safe communication channels between the communicating entities.
- Based on the aforementioned mechanisms, mutual authentication is established during TLS communication between SAEs and KMEs. Digital signatures can be incorporated into transmitted packets, and the public keys contained in the certificates of the communicating parties can be used to verify the corresponding signatures, thereby providing device identity authentication. Since the digital signatures are based on ML-DSA, the proposed mechanism provides a quantum-safe approach to mutual authentication.

The procedures for an SAE invoking the Get Key interface of a KME to obtain a Key Container object, as well as the procedures for establishing a SS between TNs through QKD, are consistent with those described in Section III.A and are therefore omitted here for brevity.

### B. *The Proposed Quantum-Safe TLS*

#### 1) *QKD-based TLS*

The QKD-based TLS protocol proposed in this study considers EE1 as the initiator and EE2 as the responder, as illustrated in Figs. 5 and 6.

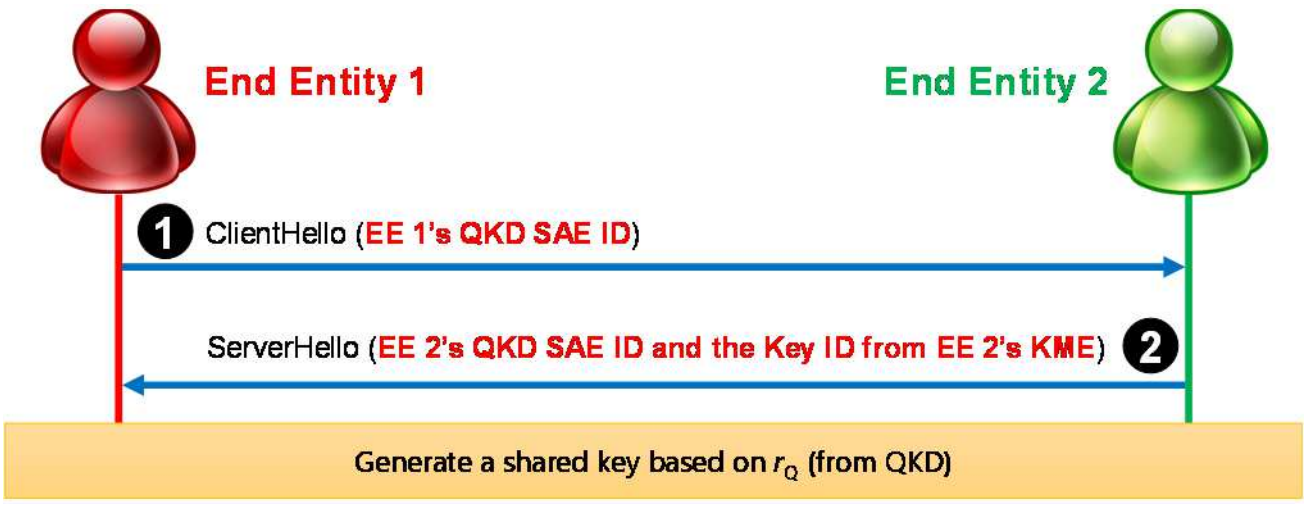

Fig. 5. The proposed QKD-based TLS

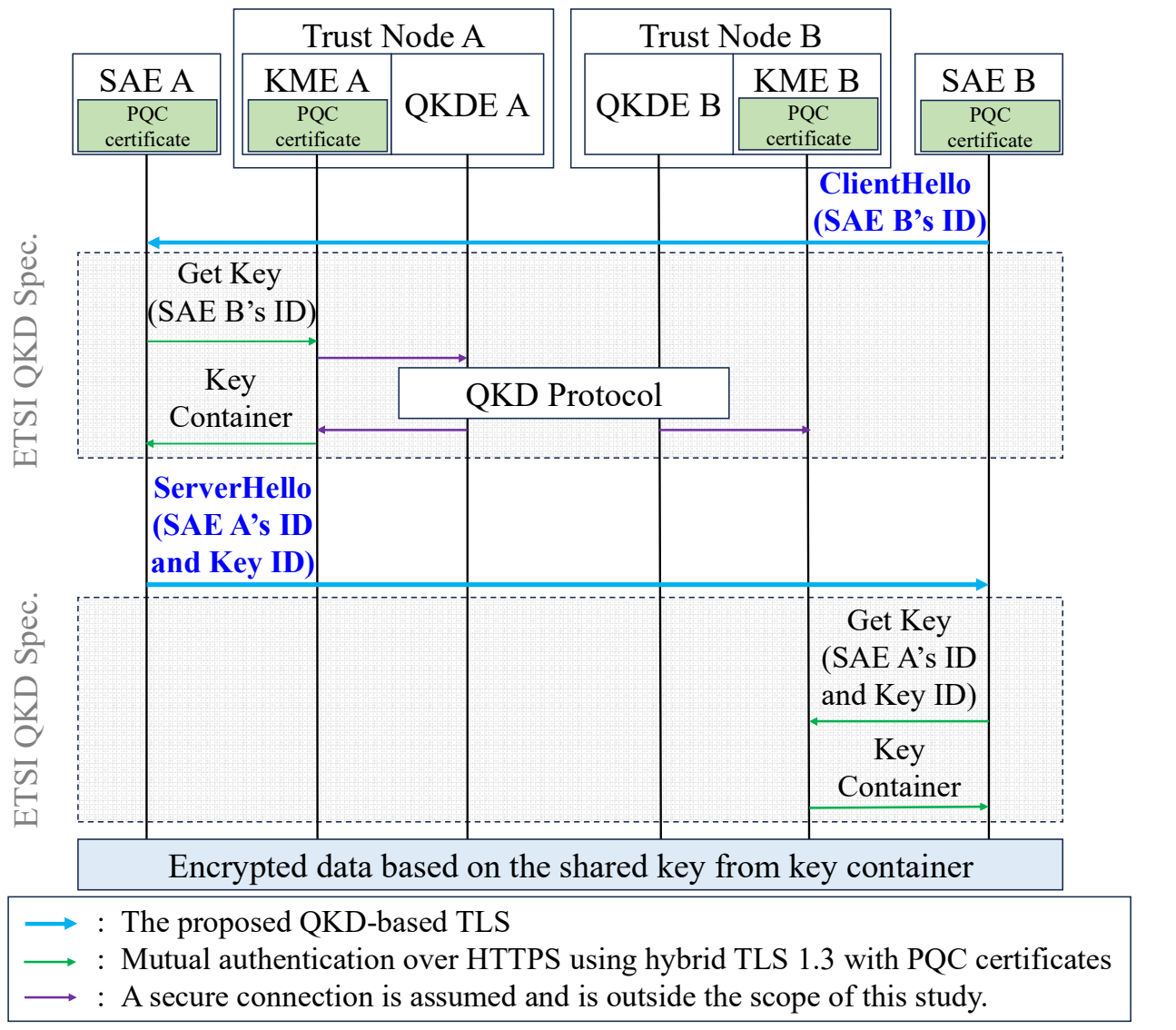

Fig. 6. The proposed QKD-based TLS under the proposed quantum-safe QKD network architecture

First, EE1 sends a ClientHello message, in which the SAE ID of EE1 is included in the key_share field. Upon receiving the ClientHello message, EE2 is regarded as SAE A in the proposed quantum-safe QKD network architecture. EE2 then calls the Get Key interface of KME A and provides the SAE ID of EE1 as the ID of SAE B. KME A subsequently returns a Key Container object containing a Key ID and the SS $r_Q$.

EE2 then sends a ServerHello message, in which the SAE ID of EE2 and the corresponding Key ID are included in the key_share field. Upon receiving the ServerHello message, EE1 is regarded as SAE B in the proposed quantum-safe QKD network architecture. EE1 then invokes the Get Key interface of KME B and provides the SAE ID of EE2 and the corresponding Key ID. KME B returns a Key Container object, from which EE1 obtains the same SS $r_Q$.

Consequently, both communicating parties obtain the SS $r_Q$. The SS is then provided as input to a KDF to derive the SK. The sender can subsequently use the SK to encrypt plaintext and transmit the resulting ciphertext, while the receiver can use the same SK to decrypt the ciphertext and recover the plaintext, thereby establishing TLS-protected communication.

*2) Hybrid QKD-PQC-based TLS*

To establish the SK based on different security properties, this study proposes a TLS protocol based on a hybrid scheme integrating QKD and PQC (named "hybrid QKD-PQC-based TLS"). The proposed approach provides security derived from both quantum physical properties and the hardness of the LWE problem. The proposed protocol is primarily based on the approaches described in Sections III.A and III.B.1, with additional information associated with the ML-KEM. In this scenario, EE1 acts as the initiator and EE2 acts as the responder, as illustrated in Fig. 7.

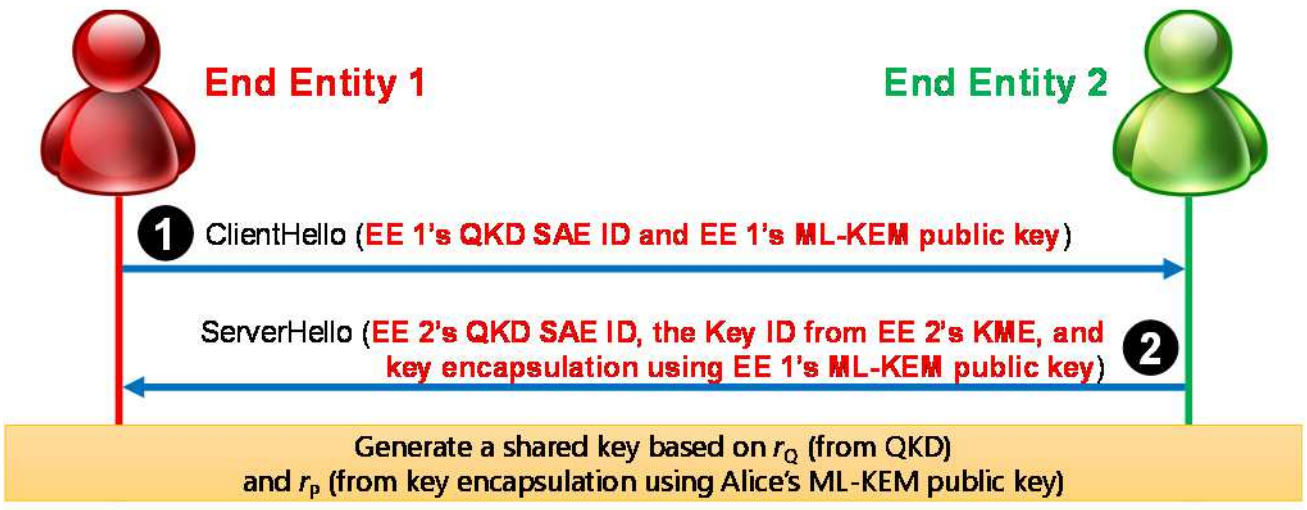

Fig. 7. The proposed hybrid QKD-PQC-based TLS

EE1 generates an ephemeral ML-KEM key pair. EE1 then sends a ClientHello message, in which the SAE ID of EE1 and the public key of the ephemeral ML-KEM key pair are included in the key_share field. Upon receiving the ClientHello message, EE2 is regarded as SAE A in the proposed quantum-safe QKD network architecture. EE2 invokes the Get Key interface of KME A and provides the SAE ID of EE1 as the ID of SAE B to obtain a Key Container object. The Key Container contains a Key ID and a SS $r_Q$. Furthermore, EE2 performs the ML-KEM encapsulation operation using EE1's ephemeral ML-KEM public key, thereby generating an ML-KEM SS $r_P$ and the corresponding encapsulation ciphertext.

EE2 then sends a ServerHello message, in which the SAE ID of EE2, the Key ID, and the ML-KEM encapsulation ciphertext are included in the key_share field. Upon receiving the ServerHello message, EE1 is regarded as SAE B in the proposed quantum-safe QKD network architecture. EE1 invokes the Get Key interface of KME B and provides the SAE ID of EE2 and the corresponding Key ID to obtain a Key Container object containing the SS $r_Q$. Furthermore, EE1 uses its ephemeral ML-KEM private key to perform the ML-KEM decapsulation operation on the encapsulation ciphertext and recovers the same ML-KEM SS $r_P$.

Finally, both communicating parties obtain the SSs $r_Q$ and $r_P$. These two SSs are provided as inputs to a KDF to derive the final SK. Consequently, the resulting SK incorporates security properties derived from both QKD and PQC.

*3) Hybrid QKD-ECC-PQC-based TLS*

Although ECC may become vulnerable to sufficiently powerful quantum computers in the future, it has been extensively studied and widely deployed for many years and remains a commonly used public-key cryptographic mechanism. Consequently, hybrid approaches that incorporate ECC alongside quantum-resistant mechanisms have also been considered. Accordingly, this study proposes a TLS protocol based on a hybrid scheme integrating QKD, ECC, and PQC (named "hybrid QKD-ECC-PQC-based TLS"). The proposed hybrid QKD-ECC-PQC-based TLS combines security properties derived from quantum physical properties, the LWE problem, and the ECDLP. The proposed protocol primarily combines the approaches described in Sections II.B.1 and III.B.2. In the considered scenario, EE1 acts as the initiator and EE2 acts as the responder, as illustrated in Fig. 8.

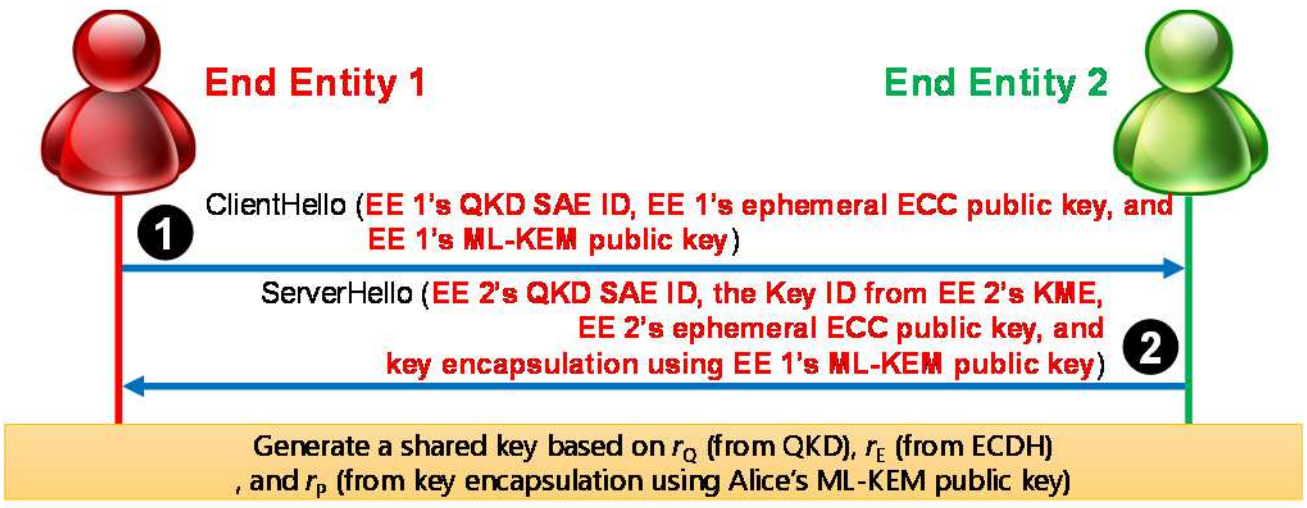

Fig. 8. The proposed hybrid QKD-ECC-PQC-based TLS

EE1 sends a ClientHello message containing the SAE ID of EE1, the ephemeral ECC public key, and the ephemeral ML-KEM public key. Subsequently, EE2 sends a ServerHello message containing the SAE ID of EE2, the Key ID, the ephemeral ML-KEM public key, and the ML-KEM encapsulation ciphertext. The detailed procedures are consistent with those described in the preceding sections and are therefore omitted for brevity.

## IV. Experimental Results and Discussions

### *A. Experimental Environments and a Case Study*

This section describes the configurations of the SAEs and KMEs and provides a case study of the experimental setup. The relevant configurations are summarized as follows:

- SAE A: The SAE ID is "CHT_Test_SAE_A", with a size of 14 bytes. SAE A is equipped with an ML-DSA-87 key pair and supports mutual authentication with KME A.
- SAE B: The SAE ID is "CHT_Test_SAE_B", with a size of 14 bytes. SAE B is equipped with an ML-DSA-87 key pair and supports mutual authentication with KME B.
- KME A: The domain name is "Test_KEM_A.qkd.hinet.net". KME A can interact with QKDE A to obtain a SS through QKD protocol.
- KME B: The domain name is "Test_KEM_B.qkd.hinet.net". KME B can interact with QKDE B to obtain a SS through QKD protocol.
- Certificate Authority (CA): The CN of the CA used in this study is "CHT_Test_RootCA". The CA uses ML-DSA-87 to issue X.509 certificates to the SAEs and KMEs.
- QKD groups: The QKD-generated SSs considered in this study have lengths of 128, 192, and 256 bits, which are designated as QKD-128, QKD-192, and QKD-256, respectively, as the corresponding group names.
- Key ID: The randomly generated Key IDs used in this study follow the Universally Unique Identifier (UUID) format and are represented as strings. For instance, "b39d9fa7-790d-4a27-95e1-59f5dfdd2957" has a size of 36 bytes.

Taking SAE A as an instance, the CN in its X.509 certificate is "CHT_Test_SAE_A", which is consistent with the SAE ID, as illustrated in Fig. 9. The certificate is issued by CHT_Test_RootCA using ML-DSA-87. Furthermore, the public-key field in the certificate contains the ML-DSA-87 public key of SAE A. This public key can subsequently be used for signature verification, thereby supporting mutual authentication between SAE A and the corresponding KME.

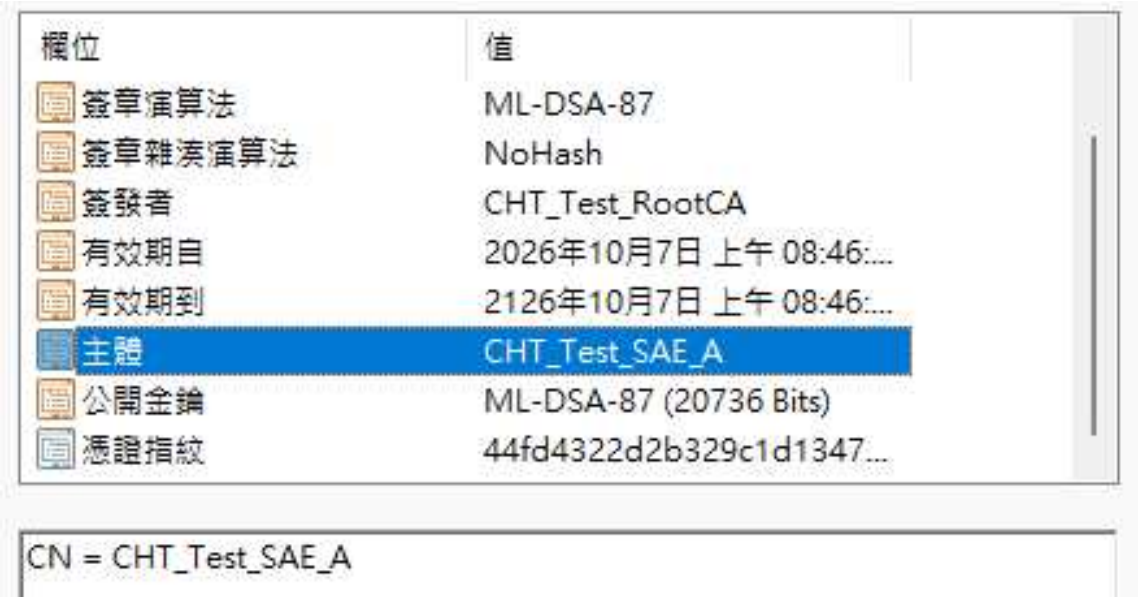

Fig. 9. The PQC-based X.509 certificate of SAE A

For a case study of QKD-256, after SAE A obtains the ID of SAE B, SAE A calls the Get Key interface of KME A at https://Test_KEM_A.qkd.hinet.net/api/v1/keys/CHT_Test_SAE_B/enc_keys and sends the request message shown in Fig. 10 to KME A. After performing the QKD negotiation procedure, KME A returns the response message shown in Fig. 11. The key field contains the SS encoded in Base64 format. After Base64 decoding, the resulting value is a 256-bit random value. Subsequently, SAE A provides the ID of SAE A and the corresponding Key ID to SAE B. SAE B then invokes the Get Key with Key IDs interface of KME B at https://Test_KEM_B.qkd.hinet.net/api/v1/keys/CHT_Test_SAE_A/dec_keys and sends the request message shown in Fig. 12 to KME B. KME B returns the response message shown in Fig. 11, from which SAE B obtains the same SS.

```
{"size": 256}
```

Fig. 10. The request message of Get Key

```
{
  "keys": [{"key_ID": "b39d9fa7-790d-4a27-95e1-59f5dfdd2957",
    "key": "DDg54Fm8isHeys2whtVXNdsmIPMErrHuJP4VqWWpHxQ="}]
}
```

Fig. 11. The response message of Get Key

```
{
  "key_IDs": [{"key_ID" : "b39d9fa7-790d-4a27-95e1-59f5dfdd2957"}]
}
```

Fig. 12. The request of Get Key with Key IDs

## B. *Security Discussions*

The proposed approaches are evaluated using different parameter combinations and compared with existing hybrid ECC-PQC-based TLS, with the results presented in Table I. The proposed hybrid QKD-ECC-PQC-based TLS derives its security from three complementary security foundations: QPP, the LWE problem, and the ECDLP. With the combination of QKD-256, NIST P-521, and ML-KEM-1024, the proposed scheme provides a security strength of up to 256 bits.

## C. *Performance Discussions*

The sizes of the key_share fields, excluding the Group ID, in the ClientHello and ServerHello messages of the TLS protocols are compared, with the results presented in Table II.

TABLE I. THE COMPARISON RESULTS OF SECURITY

| TLS | Parameter Set | Security Strength | QPP | LWE | ECDLP |
|---|---|---|---|---|---|
| ECC-based TLS [12] | X25519 | 128 | | | x |
| | NIST P-256 | 128 | | | x |
| | NIST P-384 | 192 | | | x |
| Hybrid ECC-PQC-based TLS [13],[14] | X25519, ML-KEM-768 | 128 | | x | x |
| | NIST P-256, ML-KEM-768 | 128 | | x | x |
| | NIST P-384, ML-KEM-1024 | 192 | | x | x |
| QKD-based TLS (The Proposed Method) | QKD-128 | 128 | x | | |
| | QKD-192 | 192 | x | | |
| | QKD-256 | 256 | x | | |
| Hybrid QKD-PQC-based TLS (The Proposed Method) | QKD-128, ML-KEM-768 | 128 | x | x | |
| | QKD-192, ML-KEM-768 | 192 | x | x | |
| | QKD-256, ML-KEM-1024 | 256 | x | x | |
| Hybrid QKD-ECC-PQC-based TLS (The Proposed Method) | QKD-128, X25519, ML-KEM-768 | 128 | x | x | x |
| | QKD-128, NIST P-256, ML-KEM-768 | 128 | x | x | x |
| | QKD-192, NIST P-384, ML-KEM-1024 | 192 | x | x | x |
| | QKD-256, NIST P-521, ML-KEM-1024 | 256 | x | x | x |

Since the proposed QKD-based TLS primarily transmits the SAE ID and Key ID in the ClientHello and ServerHello messages, its message size remains independent of the QKD key size. Furthermore, because the ClientHello message in the proposed QKD-based TLS only transmits the SAE ID, its size is smaller than that of the ClientHello message in the ECC-based TLS. Similarly, the ClientHello message of the proposed hybrid QKD-PQC-based TLS is smaller than that of the hybrid ECC-PQC-based TLS.

In ECC-based TLS, the public key size of X25519 is 32 bytes. Therefore, the key_share field in both the ClientHello and ServerHello messages occupies 32 bytes. For NIST P-256, the public key is represented as an uncompressed elliptic curve point and is encoded as (0x04 || 32-byte X-coordinate || 32-byte Y-coordinate), resulting in a total size of 65 bytes. Therefore, the key_share field in both the ClientHello and ServerHello messages occupies 65 bytes.

The Hybrid ECC-PQC-based TLS adopts the format (ECC public key || PQC public key or encapsulation ciphertext). For the NIST elliptic curve point representation, the leading 0x04 byte is omitted, and only the (X-coordinate || Y-coordinate) is included. Therefore, for the hybrid scheme combining NIST P-256 and ML-KEM-768, the ClientHello message contains (32-byte X-coordinate || 32-byte Y-coordinate || 1184-byte ML-KEM-768 public key), resulting in a total size of 1248 bytes. The ServerHello message contains (32-byte X-coordinate || 32-byte Y-coordinate || 1088-byte ML-KEM-768 ciphertext), resulting in a total size of 1152 bytes.

TABLE II. THE COMPARISON RESULTS OF MESSAGE SIZES (UNIT: BYTES)

| TLS | Parameter Set | ClientHello | ServerHello |
|---|---|---|---|
| ECC-based TLS [12] | X25519 | 32 | 32 |
| | NIST P-256 | 65 | 65 |
| | NIST P-384 | 97 | 97 |
| Hybrid ECC-PQC-based TLS [13],[14] | X25519, ML-KEM-768 | 1216 | 1120 |
| | NIST P-256, ML-KEM-768 | 1248 | 1152 |
| | NIST P-384, ML-KEM-1024 | 1664 | 1664 |
| QKD-based TLS (The Proposed Method) | QKD-128 | 14 | 50 |
| | QKD-192 | 14 | 50 |
| | QKD-256 | 14 | 50 |
| Hybrid QKD-PQC-based TLS (The Proposed Method) | QKD-128, ML-KEM-768 | 1198 | 1138 |
| | QKD-192, ML-KEM-768 | 1198 | 1138 |
| | QKD-256, ML-KEM-1024 | 1582 | 1618 |
| Hybrid QKD-ECC-PQC-based TLS (The Proposed Method) | QKD-128, X25519, ML-KEM-768 | 1230 | 1170 |
| | QKD-128, NIST P-256, ML-KEM-768 | 1262 | 1202 |
| | QKD-192, NIST P-384, ML-KEM-1024 | 1678 | 1714 |
| | QKD-256, NIST P-521, ML-KEM-1024 | 1714 | 1750 |

In the proposed hybrid QKD-PQC-based TLS, the ClientHello message adopts the format (SAE ID || PQC public key), while the ServerHello message adopts the format (SAE ID || Key ID || PQC encapsulation ciphertext). For the hybrid scheme combining QKD-192 and ML-KEM-768, the ClientHello message contains (14-byte SAE ID || 1184-byte ML-KEM-768 public key), resulting in a total size of 1198 bytes. The ServerHello message contains (14-byte SAE ID || 36-byte Key ID || 1088-byte ML-KEM-768 encapsulation ciphertext), resulting in a total size of 1138 bytes. The QKD-192 and ML-KEM-768 hybrid scheme has smaller message sizes than the corresponding NIST P-256 and ML-KEM-768 hybrid scheme. Furthermore, the QKD-192 and ML-KEM-768 hybrid scheme provides a security strength of up to 192 bits, whereas the NIST P-256 and ML-KEM-768 hybrid scheme provides a security strength of approximately 128 bits.

## V. CONCLUSIONS AND FUTURE WORK

This study proposes a quantum-safe QKD network architecture and quantum-safe TLS protocols. Experimental results demonstrate that the proposed approaches can provide a high level of security while maintaining relatively small message sizes. Future research may focus on further investigating the communication between KMEs and QKDEs, with the aim of improving both security and performance.

## ACKNOWLEDGMENT

The author would like to express their sincere appreciation to the ETSI QKD Working Group (WG) for considering and incorporating the relevant comments into the GSs, including ETSI GS QKD 014 and ETSI GS QKD 020, which specify the interfaces between the SAE and KME, and between KMEs, respectively. The author also gratefully acknowledges the IETF for considering the relevant comments and incorporating them into the standards document RFC 9882, which specifies the Cryptographic Message Syntax (CMS) for ML-DSA digital signatures. Moreover, the author sincerely thanks the IETF PQC Certificate WG for considering the relevant feedback and conducting interoperability testing.